\documentclass[aps,twocolumn,superscriptaddress,nofootinbib]{revtex4-1} 
\usepackage{dcolumn}    
\usepackage{pstricks}
\usepackage{color}
\usepackage{graphicx}
\usepackage{amsmath}
\usepackage{mathtools}
\usepackage[T1]{fontenc}
\usepackage[utf8]{inputenc}
\usepackage{ae}
\usepackage{amssymb}
\usepackage[colorlinks=true,allcolors=darkred, pdfborder={0 0 0}]{hyperref}

\usepackage{booktabs}

\def\bqn#1{\begin{equation}\label{eq:#1}}
\def\eqn{\end{equation}}

\def\bqa#1{\begin{align}\label{eq:#1}}
\def\eqa{\end{align}}

\newcommand{\rep}[1]{\mathbf{#1}}
\newcommand{\repb}[1]{\mathbf{\overline{#1}}}

\definecolor{darkred}{rgb}{0.6,0,0}

\begin{document}

\title{A minimal non-supersymmetric SO(10) model with Peccei--Quinn symmetry}

\author{Sofiane M.~Boucenna}
\email{boucenna@kth.se}
\affiliation{Department of Physics,
	School of Engineering Sciences,
	KTH Royal Institute of Technology,
	AlbaNova University Center,
	Roslagstullsbacken 21,
	SE--106 91 Stockholm,
	Sweden}
\affiliation{The Oskar Klein Centre for Cosmoparticle Physics,
	AlbaNova University Center,
	Roslagstullsbacken 21,
	SE--106 91 Stockholm,
	Sweden}

\author{Tommy Ohlsson}
\email{tohlsson@kth.se}
\affiliation{Department of Physics,
	School of Engineering Sciences,
	KTH Royal Institute of Technology,
	AlbaNova University Center,
	Roslagstullsbacken 21,
	SE--106 91 Stockholm,
	Sweden}
\affiliation{The Oskar Klein Centre for Cosmoparticle Physics,
	AlbaNova University Center,
	Roslagstullsbacken 21,
	SE--106 91 Stockholm,
	Sweden}
\affiliation{University of Iceland, 
	Science Institute, 
	Dunhaga 3, 
	IS--107 Reykjavik, 
	Iceland}

\author{Marcus Pernow}
\email{pernow@kth.se}
\affiliation{Department of Physics,
	School of Engineering Sciences,
	KTH Royal Institute of Technology,
	AlbaNova University Center,
	Roslagstullsbacken 21,
	SE--106 91 Stockholm,
	Sweden}
\affiliation{The Oskar Klein Centre for Cosmoparticle Physics,
	AlbaNova University Center,
	Roslagstullsbacken 21,
	SE--106 91 Stockholm,
	Sweden}

\begin{abstract}
We present a minimal non-supersymmetric SO(10) GUT breaking directly to the Standard Model gauge group. Precise gauge coupling unification is achieved due to the presence of two color-octet scalars, one of which is accessible to LHC searches. Proton lifetime is predicted to be below $4.5\times 10^{34}$ years, which is within the projected five-year sensitivity of the proposed Hyper-Kamiokande experiment. We find that the Standard Model observables are reproduced to a reasonable accuracy in a numerical fit, which also predicts the unknown neutrino parameters. Finally, the two scalar representations stabilize the electroweak vacuum and the dark matter is comprised of axions.
\end{abstract}

\maketitle

\section{Introduction}

Non-supersymmetric SO(10) grand unified theories (GUTs) \cite{Georgi:1975qb,Fritzsch:1974nn} provide an appealing framework for physics beyond the Standard Model (SM). In addition to providing a unified description of the SM gauge group, they can naturally describe neutrino masses and the baryon asymmetry \cite{Fong:2014gea}, provide a dark matter candidate in the form of axions \cite{Reiss:1981nd,Mohapatra:1982tc,Holman:1982tb,Ernst:2018bib} or weakly interacting massive particles \cite{Frigerio:2009wf,Kadastik:2009cu,Boucenna:2015sdg}. 

In this letter, we will present an SO(10) GUT, which addresses the unification of the SM  and the problem of dark matter in a minimal way. Contrary to  common practice, we will not invoke an intermediate breaking step between the electroweak scale and the GUT scale (compare e.g.~Refs.~\cite{Bertolini:2009qj,Altarelli:2013aqa,Babu:2015bna}). Indeed, we will break SO(10) directly to the SM gauge group.\footnote{For previous attempts at constructing non-supersymmetric SO(10) GUTs with direct breaking to the SM, see e.g.~Refs.~\cite{Frigerio:2009wf,Parida:2016hln}.} Fermion observables are obtained via the actions of the $\rep{10}+\repb{126}$ Higgs representations, which have been previously shown to be viable \cite{Dueck:2013gca} and contain the necessary ingredients to implement leptogenesis. We will consider a global Peccei--Quinn (PQ) symmetry \cite{Peccei:1977ur} to solve the strong CP problem and provide a dark matter candidate \cite{Peccei:1977ur,Peccei:1977hh,Weinberg:1977ma,Wilczek:1977pj}. The PQ symmetry is broken at the GUT scale for the sake of minimality, since the lack of intermediate steps means that there is a priori no reason for the PQ fields to take vacuum expectation values (vevs) at any other scale.

This letter is organized as follows: First, in Sec.~\ref{sec:model}, we describe our model and outline its most salient features. Then, in Sec.~\ref{sec:gcu}, we analyze the constraints on gauge coupling unification and the predictions for proton lifetime. In Sec.~\ref{sec:fits}, we perform global fits to the parameters of our model and analyze the results and predictions of the model. Next, in Sec.~\ref{sec:axions}, we briefly discuss how axion dark matter fits in our model and implications on inflation. Finally, in Sec.~\ref{sec:conclusions}, we summarize and present our conclusions.

\section{Description of the Model}
\label{sec:model}

We consider a GUT  based on the symmetry $\mathrm{SO}(10)\times \mathrm{U}(1)_{\mathrm{PQ}}$, where $\mathrm{U}(1)_{\mathrm{PQ}}$ is the global PQ symmetry. The fermions are in the spinorial  $\rep{16}_F$ representation and the Yukawa interactions are obtained via the complexified $\rep{10}_H$ and $\repb{126}_H$ Higgs representations. The breaking of  $\mathrm{SO}(10)\times \mathrm{U}(1)_{\mathrm{PQ}}$ is achieved using the $\rep{210}_H$, $\repb{126}_H$, and $\rep{45}_H$ representations and proceeds in one step, directly to the SM symmetry. The electroweak gauge group is broken via the $\mathrm{SU}(2)_L$ doublets in the $\rep{10}_H$ and $\repb{126}_H$ representations. Schematically, it follows that
\begin{align}
\mathrm{SO}(10)\times \mathrm{U}(1)_{\mathrm{PQ}} &\longrightarrow \mathrm{SU}(3)_C\times \mathrm{SU}(2)_L \times \mathrm{U}(1)_Y \nonumber\\
&\longrightarrow \mathrm{SU}(3)_C\times \mathrm{U}(1)_Q \,.
\label{eq:breaking}
\end{align}
The PQ charges are such that
\begin{align}
&\rep{16}_F \to e^{{\rm i}\alpha}\rep{16}_F \,, \nonumber\\
& \rep{10}_H \to e^{-2{\rm i}\alpha}\rep{10}_H\,,\quad \repb{126}_H \to e^{-2{\rm i}\alpha}\repb{126}_H\,, \nonumber\\
& \rep{210}_H \to  \rep{210}_H\,,\quad  \rep{45}_H \to e^{4{\rm i}\alpha} \rep{45}_H\,,
\label{eq:pqcharges}
\end{align}
where $\alpha$ is a real number, and the Lagrangian of the Yukawa interactions reads
\begin{equation}
\mathcal{L}_Y= \rep{16}_F (Y_{10} \rep{10}_H + Y_{126} \repb{126}_H) \rep{16}_F\,,
\label{eq:yakuawa}
\end{equation}
where the Yukawa couplings $Y_{10}$ and $Y_{126}$ are $3 \times 3$ matrices in flavor space. Although the $\rep{10}_H$ is complexified to allow for two different vevs (see below) as phenomenologically required, the PQ symmetry forbids the Yukawa interactions with $\rep{10}_H^*$, thus retaining minimality of the model~\cite{Holman:1982tb,Bajc:2005zf}. After the breaking of SO(10), the Yukawa couplings of the SM are formed by combinations of $Y_{10}$ and $Y_{126}$:
\begin{align}
Y_u&=\bar{v}^u_{10} Y_{10} + \bar{v}^u_{126}Y_{126}\,, \label{eq:matching1}\\
Y_d&=\bar{v}^d_{10} Y_{10} + \bar{v}^d_{126}Y_{126} \,, \label{eq:matching2}\\
Y_\nu &=\bar{v}^u_{10} Y_{10} - 3\bar{v}^u_{126}Y_{126}\,, \label{eq:matching3}\\  
Y_\ell &=\bar{v}^d_{10} Y_{10} - 3\bar{v}^d_{126}Y_{126}\,, \label{eq:matching4}
\end{align}
where $Y_u$, $Y_d$, $Y_\nu$, and $Y_\ell$ are Yukawa couplings for the up-type quarks, down-type quarks, neutrinos, and charged leptons, respectively. The quantities $v^{u}_{10}$, $v^{d}_{10}$, $v^{u}_{126}$, and $v^{d}_{126}$ are the vevs of the $\mathrm{SU}(2)_L$ Higgs doublets in the $\rep{10}_H$ and $\repb{126}_H$, respectively, and we define $\bar{x}\equiv x/{v_\mathrm{SM}}$ with $v_\mathrm{SM} \simeq 174.1\,\mathrm{GeV}$. These vevs satisfy $(\bar{v}^u_{10})^2 + (\bar{v}^d_{10})^2 + (\bar{v}^u_{126})^2 + (\bar{v}^d_{126})^2 = 1$, and we assume that only one physical combination of these $\mathrm{SU}(2)_L$ Higgs doublets survives at low energy, corresponding to the SM Higgs doublet. Neutrino masses are generated using the seesaw mechanism \cite{Minkowski:1977sc,GellMann:1980vs,yanagida:1979,Mohapatra:1979ia,Lazarides:1980nt,Schechter:1980gr} with the right-handed neutrinos obtaining masses from the SM singlet $\sigma$ contained in the $\repb{126}_H$, namely
\begin{equation}\label{eq:matching5}
M_R=v^\sigma Y_{126} \,.
\end{equation}

\section{Gauge Coupling Unification and Proton Decay}
\label{sec:gcu}

Precise gauge coupling unification can be achieved in our model after direct breaking of SO(10) to the SM due to the presence of two color-octet scalar multiplets from the $\mathbf{210}_H$ representation used in the breaking, namely $S_1\equiv(\mathbf{8},\mathbf{1},1)$ and $S_2\equiv(\mathbf{8},\mathbf{3},0)$. We assume the existence of a fine-tuning of the parameters of the scalar potential  leading to the required splitting of the $\mathbf{210}_H$.\footnote{Note that this may require introducing a new scalar representation to the model in order to have extra couplings. However, these new representations will be integrated out at the GUT scale and will have no impact on the model other than facilitating the splitting of the $\mathbf{210}_H$.}  Similar constructions of extra multiplets surviving at a lower scale to facilitate gauge coupling unification have been previously studied in $\mathrm{SU}(5)$-based models in e.g.~Refs.~\cite{Bajc:2006ia,Bajc:2007zf,DiLuzio:2013dda,Boucenna:2017fna}. The representations $S_{1}$ and $S_{2}$ with masses $M_{1}$ and $M_{2}$, respectively, satisfying $M_1 \leq M_2 \leq M_\mathrm{GUT}$, alter the renormalization group (RG) running such that gauge coupling unification is obtained if
\begin{align}
M_1&\lesssim 5.92\times 10^{10} \,  \mathrm{GeV} \label{eq:MGUT1}\,, \\
M_2&\approx\left(\frac{M_1}{\mathrm{GeV}}\right)^{0.330}\times1.65\times 10^7 \, \mathrm{GeV}\,, \\
M_\mathrm{GUT}&\approx\left(\frac{M_1}{\mathrm{GeV}}\right)^{-0.0447}\times7.34\times 10^{15} \, \mathrm{GeV}\,. 
\label{eq:MGUT}
\end{align}
Equations~\eqref{eq:MGUT1}--\eqref{eq:MGUT} were obtained by solving the renormalization group equations (RGEs) at one-loop order, neglecting threshold effects.

The proton lifetime $\tau_p$ is directly related to the GUT scale,  $M_\mathrm{GUT}$. The most constraining decay channel is via the dimension-six operator for $p\to e^+\pi^0$ and an approximate relation for $\tau_p$ is given by \cite{Babu:2010ej,Sahoo:2017cqg}
\begin{equation}
\tau_p\equiv\tau(p\to e^+\pi^0)\simeq \frac{4}{\pi} \frac{f_\pi^2}{m_p}\frac{1}{\alpha_H^2 A_R^2}\frac{1}{F_q}\frac{M_\mathrm{GUT}^4}{\alpha(M_\mathrm{GUT})^2} \,,
\label{eq:tau_p}
\end{equation}
where $f_\pi \approx 139\,\mathrm{MeV}$ is the pion decay constant, $m_p \approx 938.3\,\mathrm{MeV}$ is the proton mass, $\alpha_H \approx 0.012\,\mathrm{GeV}^3$ is the hadronic matrix element, $A_R \approx 2.726$ is a renormalization factor, and $F_q \approx 7.6$ is a quark-mixing factor. Using these input parameter values, Eq.~(\ref{eq:tau_p}) simplifies to
\begin{equation}
\tau_p \approx (3.22\,\text{GeV}^{-5}) \times \frac{M_\mathrm{GUT}^4}{\alpha(M_\mathrm{GUT})^2}\,.
\end{equation}

\begin{figure}
\includegraphics[clip,width=0.45\textwidth]{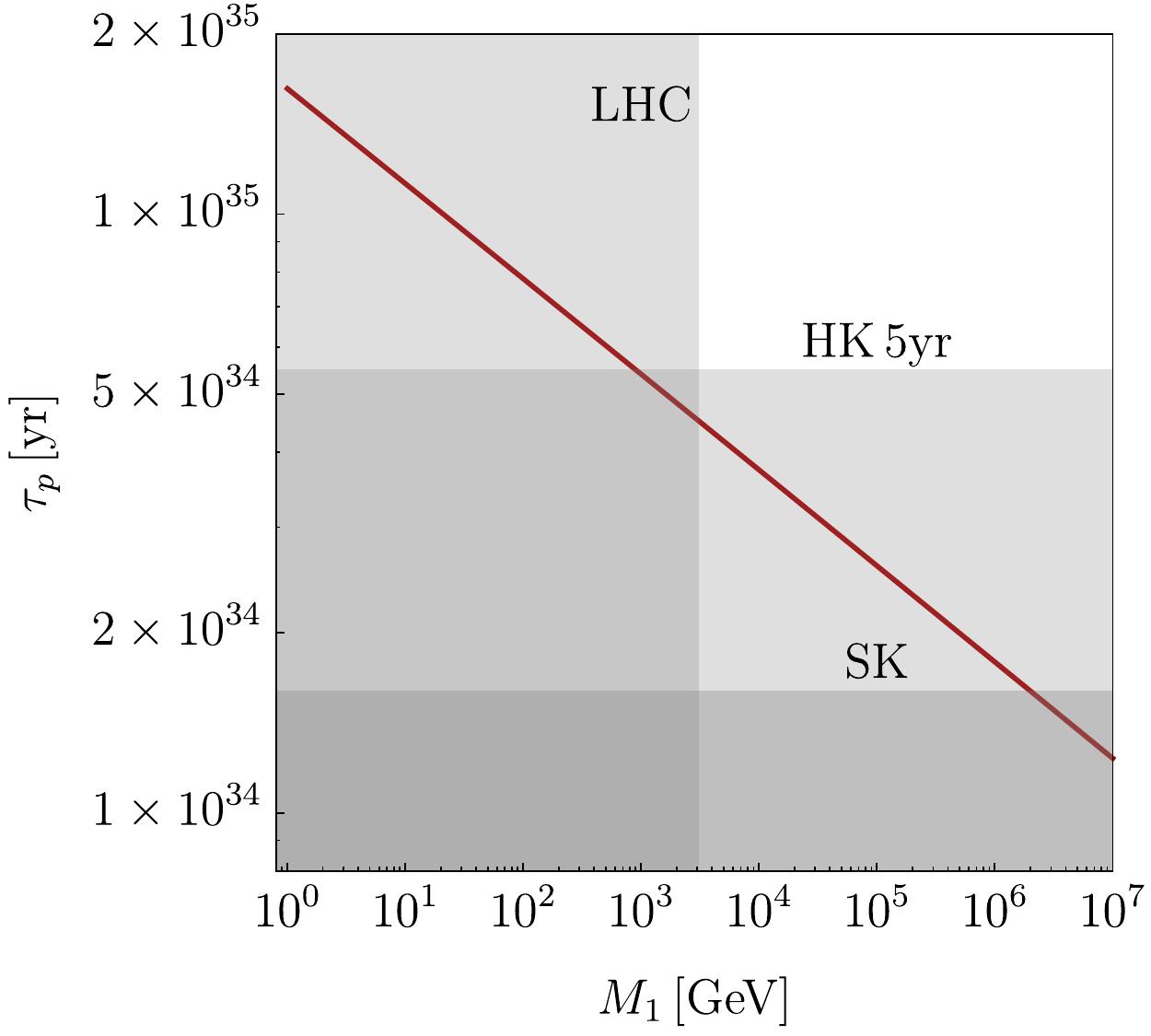}
\caption{\label{fig:proton}Variation of proton lifetime $\tau_p$ with mass $M_1$. The horizontal boundaries correspond to constraints on $\tau_p$ from Super-Kamiokande (SK) and predicted five-year constraint from Hyper-Kamiokande (HK 5yr). The vertical boundary shows LHC constraints on the mass of color-octet scalars.}
\end{figure}

In Fig.~\ref{fig:proton}, we display the relationship between $M_1$ and $\tau_p$  together with the current bound from Super-Kamiokande, $\tau_p > 1.6\times 10^{34} \,\mathrm{yr}$ \cite{Miura:2016krn,Tanabashi:2018oca}, the projected five-year discovery potential of Hyper-Kamiokande, $\tau_p > 5.5\times 10^{34} \,\mathrm{yr}$ \cite{Abe:2018uyc}, as well as the current LHC lower bound  on the mass of the color-octet $S_1$, $M_1>3.1\,\mathrm{TeV}$ \cite{Khachatryan:2015dcf}.\footnote{Note that the color-octet representation tested in Ref.~\cite{Khachatryan:2015dcf}  is not the same as ours. However, a similar bound is expected to hold in our case.} The current constraint on $\tau_p$ from Super-Kamiokande gives the bound $M_1 \lesssim 1.99 \times 10^6\,\mathrm{GeV}$. Note that if Hyper-Kamiokande does not observe proton decay after five years, this would, together with the bound from LHC, strongly disfavor our model.

\begin{figure}
\includegraphics[clip,width=0.45\textwidth]{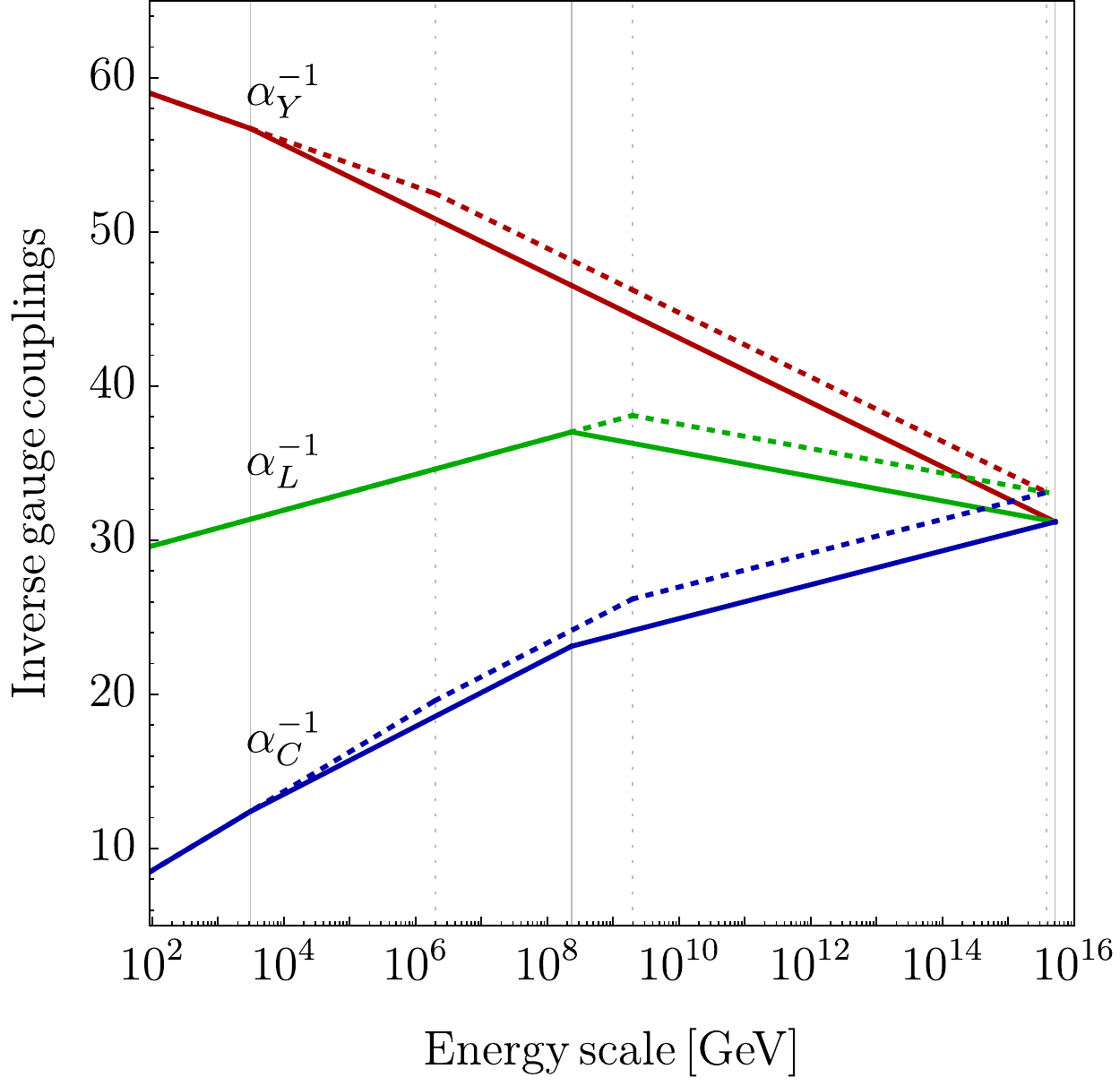}
\caption{\label{fig:gcu}Renormalization group running of the inverse gauge couplings, showing unification. The solid lines show the solution with $M_1=3.10 \times 10^3 \,\mathrm{GeV}$, $M_2=2.34\times 10^8 \,\mathrm{GeV}$, and $M_\mathrm{GUT}=4.51\times 10^{15}\,\mathrm{GeV}$, whereas the dashed lines show the solution with $M_1=1.99\times 10^6\,\mathrm{GeV}$, $M_2=1.97\times 10^9 \,\mathrm{GeV}$, and $M_\mathrm{GUT}=3.84\times 10^{15}\,\mathrm{GeV}$.}
\end{figure}

Figure~\ref{fig:gcu} shows the evolution of the inverse gauge couplings with energy for the extreme cases $M_1=3.10  \times 10^3 \,\mathrm{GeV}$ and $M_1=1.99\times 10^6\,\mathrm{GeV}$.  Within these bounds, the variation of $M_\mathrm{GUT}$ and the gauge couplings at $M_\mathrm{GUT}$ is shown to be small, which is also evident from the small exponent in Eq.~\eqref{eq:MGUT}. Finally, note that there are several other possible choices of sub-representations of the $\mathbf{210}_H$, which could also provide gauge coupling unification. We motivate our choice by selecting the minimal combination giving the lowest possible $\tau_p$, thus being directly testable.

\section{Numerical Fit of SM Observables}
\label{sec:fits}

\subsection{Numerical Procedure}

To make sure that the SM can be correctly reproduced in our model, we fit its 19 observables (5 lepton mass parameters, 6 quark masses, 3 leptonic mixing angles, 4 quark mixing parameters, and the Higgs mass) to the input parameters in Eqs.~(\ref{eq:matching1})--(\ref{eq:matching5}). The latter are conveniently parametrized as \cite{Dutta:2004zh,Dutta:2005ni,Altarelli:2010at,Joshipura:2011nn,Altarelli:2013aqa,Dueck:2013gca,Babu:2015bna}
\begin{align}
Y_u&=r(H+sF) \,, \label{eq:parametrisation1}\\
Y_d&=H+F \,, \label{eq:parametrisation2}\\
Y_\nu &=r(H-3sF)\,, \label{eq:parametrisation3}\\
Y_\ell &=H-3F \,, \label{eq:parametrisation4} \\
M_R&=t F \,, \label{eq:parametrisation5}
\end{align}
where $H\equiv  \bar{v}^d_{10}  Y_{10}$, $F\equiv \bar{v}^d_{126} Y_{126}$, $r\equiv v^u_{10}/v^d_{10}$, $s\equiv v^u_{126}/(r v^d_{126})$, and $t \equiv v^\sigma/\bar{v}^d_{126}$. The notation of $\bar{x}\equiv x/v_\mathrm{SM}$ is the same as the one introduced in Eqs.~\eqref{eq:matching1}--\eqref{eq:matching4}. This parametrization of the vevs in ratios $r$ and $s$ gives enough freedom to satisfy the electroweak constraints.
The 19 free parameters are then: 3 in $H$ (after choosing a basis in which $H$ is real and diagonal), 12 in $F$ (complex symmetric), 1 in $r$ (real), 2 in $s$ (complex), and 1 in $t$ (real). Since we have a one-step breaking, we also require $v^\sigma\sim M_\mathrm{GUT}$, corresponding to $t > M_\mathrm{GUT}$. However, since we neglect threshold effects~\cite{Weinberg:1980wa,Hall:1980kf} and higher-dimensional operators~\cite{Ellis:1979fg,Hill:1983xh,Shafi:1983gz} which may impact this relation, we allow for a slightly broader variation as $t> 0.1 M_\mathrm{GUT}$. In principle, one should also sample the value of the Higgs quartic coupling $\lambda$ at $M_\mathrm{GUT}$. However, it was observed to be consistently close to zero and small variations had a negligible effect on the predictions of the fit. Therefore, it was set to zero and not sampled. After obtaining the best-fit, we adjust the value of $\lambda$ to ensure the stability of the vacuum, as explained in Sec.~\ref{sec:results}. The final $\chi^2$ value and predictions are calculated with the adjusted value of $\lambda$.

The input data at $M_Z$ used in the fits (see Tab.~\ref{tab:data}) are obtained from the following sources: The values of the quark and charged lepton masses are taken from Tab.~3 of Ref.~\cite{Deppisch:2018flu} and $\lambda$ is computed from the values of the Higgs mass and vev therein. The quark-mixing parameters are computed from the ICHEP 2016 update by the CKMFitter Group \cite{Charles:2004jd}. For the leptonic mixing angles and neutrino mass-squared differences, we use Ref.~\cite{deSalas:2017kay} for both normal and inverted neutrino mass ordering. Similarly to  Ref.~\cite{Dueck:2013gca}, in order to improve the efficiency of the numerically challenging fits, we artificially enlarge the errors known to better than 5~\% to a minimum of 5~\% deviation from the central values.

 \begin{table}[!ht]
 \begin{center}
 \begin{tabular}{l l l l}
 \hline 
 \hline
 
 Parameter														& Central value & Error 	\tabularnewline
 \hline
 $m_u\,\mathrm{(MeV)}$								&  $1.36$	&  $0.15$ 	\tabularnewline
 $m_c\,\mathrm{(MeV)}$								&  $635$		& $32$ 		 \tabularnewline
 $m_t\,\mathrm{(GeV)}$									&  $172$		& $8.7$ \tabularnewline
 $m_d\,\mathrm{(MeV)}$								&  $2.90$	& $0.15$		 \tabularnewline
 $m_s\,\mathrm{(MeV)}$								&  $54.1$	& $2.8$ 		 \tabularnewline
 $m_b\,\mathrm{(GeV)}$								&  $2.87$	& $0.15$		 \tabularnewline
 
 $m_e\,\mathrm{(MeV)}$								&  $0.487$	& $0.025$ 	\tabularnewline
 $m_\mu\,\mathrm{(MeV)}$							&  $103$		& $5.2$ 	 	\tabularnewline
 $m_\tau\,\mathrm{(GeV)}$							&  $1.75$	& $0.088$ 	\tabularnewline
 
 $\Delta m_{21}^2\,(10^{-5}\mathrm{eV^2})$	&  $7.55$	&  $0.38$		\tabularnewline
 $\Delta m_{31}^2\,(10^{-3}\mathrm{eV^2})$ (NO)	&  $2.50$	&  $0.13$ 	\tabularnewline
 $\Delta m_{32}^2\,(10^{-3}\mathrm{eV^2})$ (IO)	&  $-2.42$	&  $0.13$ 	\tabularnewline
 
 $\sin \theta^q_{12}$										&  $0.225$& $0.012$ 	  \tabularnewline
 $\sin \theta^q_{13}$										&  $0.00372$&$0.00019$  \tabularnewline
 $\sin \theta^q_{23}$										&  $0.0418$&$0.0021$  \tabularnewline
 $\delta^q_\mathrm{CP}$												&  $1.14$&$0.058$  		  \tabularnewline
 
 $\sin^2 \theta^\ell_{12}$								&  $0.320$	&  $0.020$	 \tabularnewline
 $\sin^2 \theta^\ell_{13}$	(NO)						&  $0.0216$&  $0.0011$\tabularnewline
 $\sin^2 \theta^\ell_{13}$	(IO)						&  $0.0222$&  $0.0012$\tabularnewline
 $\sin^2 \theta^\ell_{23}$	(NO)						&  $0.547$	&  $0.030$ 	\tabularnewline
 $\sin^2 \theta^\ell_{23}$	(IO)						&  $0.551$	&  $0.030$ 	\tabularnewline
 
 $\lambda$														&  $0.516$	&  $0.026$ 		\tabularnewline

 \hline 
 \hline
 \end{tabular}
 \caption{\label{tab:data}Input data at $M_Z$ used in the fits. Details and references can be found in the main text.}
 \end{center}
 \end{table}

To fit the SO(10) parameters to the SM observables, we apply the following procedure: The 19 free parameters are sampled and transformed to the SM Yukawa and right-handed neutrino mass matrices, using Eqs.~(\ref{eq:parametrisation1})--(\ref{eq:parametrisation5}). Next, they are evolved down from $M_\mathrm{GUT}$ to the electroweak scale $M_Z$, using the RGEs at one-loop order \cite{Machacek:1983tz,Machacek:1983fi,Machacek:1984zw,Jones:1981we}. The right-handed neutrinos are integrated out at each of their respective mass scales, resulting in the effective dimension-five operator for neutrino masses \cite{Antusch:2002rr,Antusch:2005gp} which we then also run down to $M_Z$. The RGEs for the gauge couplings are properly modified at each mass scale $M_1$ and $M_2$ of the color-octet scalars $S_1$ and $S_2$. In order to gain in predictivity, we assume that the couplings between the scalars $S_1$ and $S_2$ and the Higgs play a negligible role in the running of $\lambda$. The RG running of the Higgs boson coupling is therefore dominated by the gauge, the top Yukawa, and the neutrino Yukawa couplings. At $M_Z$, the observables of the SM are calculated and compared to data using a standard $\chi^2$ estimator  (note that due to the non-linearity of the problem, it is non-trivial to interpret the $\chi^2$ function statistically, as noted, e.g., in Refs.~\cite{Bjorkeroth:2017ybg,Deppisch:2018flu}. Instead, it should be interpreted only as an indication of how easy it is to fit the data and is most useful as a comparison).

In order to minimize the $\chi^2$ function, we link the code performing the procedure described above to the sampler \texttt{Diver} from the \texttt{ScannerBit} package \cite{Workgroup:2017htr}. The best-fit parameter values returned from this program are then further improved using the basin-hopping algorithm \cite{bh1997} from the \texttt{Scipy} library \cite{scipy2001}. To maximize the chance that the actual best-fit parameter values are found, we repeat this procedure multiple times.

\subsection{Results and Predictions}\label{sec:results}

For normal neutrino mass ordering, we find that the best-fit parameter values result in $\chi^2 \simeq 21.0$. The predicted values and pulls are displayed in Tab.~\ref{tab:results}. The largest contribution to the $\chi^2$ function originates from the leptonic mixing parameter $\sin^2\theta_{23}^\ell$, for which we obtain $0.445$, which is in the first octant, compared to the $1\sigma$ range $(0.517,0.577)$ which lies in the second octant. However, note that the octant of $\theta_{23}^\ell$ is still largely uncertain and values in the lower octant are still allowed by global neutrino oscillation fits~\cite{deSalas:2017kay}. Such a tension has also been observed in previous fits \cite{Dueck:2013gca}. The other two contributions to the $\chi^2$ function that are larger than unity stem from the down-quark mass $m_d$, which is found to be $2.70\,\mathrm{MeV}$, while the 5~\% range is $(2.75,3.05)\,\mathrm{MeV}$, and the muon mass $m_\mu$, which is found to be $0.111\,\mathrm{GeV}$, while the 5~\% range is $(0.0978,0.108)\,\mathrm{GeV}$. The best-fit parameter values for normal ordering are determined to be
\begin{widetext}
\begin{align}
H &= 
\begin{pmatrix}
-9.07178\times 10^{-7} && 0 && 0 \\
0 && -6.85850\times 10^{-5} && 0 \\
0 && 0 && -6.51835 \times 10^{-3}
\end{pmatrix}\,,\\
F&=
\begin{psmallmatrix}
4.46924\times 10^{-6} + 3.11197\times 10^{-16}\mathrm{i} && 5.65215\times 10^{-6}  +8.15408\times 10^{-6}\mathrm{i}	&& -1.26725\times 10^{-4} -1.39399\times 10^{-5}\mathrm{i} \\
5.65215\times 10^{-6}  +8.15408\times 10^{-6}\mathrm{i} && 1.52388\times 10^{-4} +6.77586\times 10^{-5}\mathrm{i} && -1.79155\times 10^{-4} - 3.35227\times 10^{-4}\mathrm{i} \\
-1.26725\times 10^{-4} -1.39399\times 10^{-5}\mathrm{i} && -1.79155\times 10^{-4} - 3.35227\times 10^{-4}\mathrm{i} && 8.32359\times 10^{-4} + 2.83292\times 10^{-4}\mathrm{i}
\end{psmallmatrix}\,,\\
r &= 59.8611, \quad s = 0.379855 -0.0631434\mathrm{i}, \quad t = 2.02909\times 10^{15}\,\mathrm{GeV}.
\label{eq:params}
\end{align}
\end{widetext}

 \begin{table}[!ht]
 \begin{center}
 \begin{tabular}{l l l l}

 \hline 
 \hline

 Parameter														& Predicted value & Pull 	\tabularnewline
 \hline
 $m_u\,\mathrm{(MeV)}$								&  $1.36$	&  $0.020$ 	\tabularnewline
 $m_c\,\mathrm{(MeV)}$								&  $646$		& $0.36$ 		 \tabularnewline
 $m_t\,\mathrm{(GeV)}$								&  $164$		& $-0.95$ \tabularnewline
 $m_d\,\mathrm{(MeV)}$								&  $2.70$	& $-1.4$		 \tabularnewline
 $m_s\,\mathrm{(MeV)}$								&  $54.2$	& $0.041$ 		 \tabularnewline
 $m_b\,\mathrm{(GeV)}$								&  $3.02$	& $0.99$		 \tabularnewline
 
 $m_e\,\mathrm{(MeV)}$								&  $0.505$	& $0.72$ 	\tabularnewline
 $m_\mu\,\mathrm{(MeV)}$							&  $111$		& $1.5$ 	 	\tabularnewline
 $m_\tau\,\mathrm{(GeV)}$							&  $1.76$	& $0.10$ 	\tabularnewline
 
 $\Delta m_{21}^2\,(10^{-5}\mathrm{eV^2})$	&  $7.86$	&  $0.82$		\tabularnewline
 $\Delta m_{31}^2\,(10^{-3}\mathrm{eV^2})$ &  $2.43$	&  $-0.54$ 	\tabularnewline
 
 $\sin \theta^q_{12}$										&  $0.234$& $0.75$ 	  \tabularnewline
 $\sin \theta^q_{13}$										&  $0.00373$&$0.056$  \tabularnewline
 $\sin \theta^q_{23}$										&  $0.0411$&$-0.31$  \tabularnewline
 $\delta^q_\mathrm{CP}$								&  $1.13$&$-0.23$  		  \tabularnewline
 
 $\sin^2 \theta^\ell_{12}$								&  $0.340$	&  $0.99$	 \tabularnewline
 $\sin^2 \theta^\ell_{13}$								&  $0.0215$&  $-0.12$\tabularnewline
 $\sin^2 \theta^\ell_{23}$								&  $0.445$	&  $-3.4$ 	\tabularnewline

 $\lambda$													&  $0.512$	&  $-0.14$ 		\tabularnewline

 \hline 
 \hline
 
 \end{tabular}

 \caption{\label{tab:results} Predicted values and pulls corresponding to the best-fit parameter values.}
 \end{center}
 \end{table}

The best-fit parameter values  allow us to make  predictions on the unknown neutrino parameters: (i) The sum of the light neutrino masses is $\Sigma m_\nu \simeq 6.25\times10^{-2}\,\mathrm{eV}$, which is below the upper limit from cosmological observations \cite{Tanabashi:2018oca}, (ii) the effective double beta-decay neutrino mass is predicted to be very difficult to observe \cite{Deppisch:2014zta,Pas:2015eia} with $m_{ee} \simeq 3.33\times10^{-3}$ eV, (iii) the leptonic CP-violating phase $\delta_\mathrm{CP}^\ell \simeq 0.187$ is smaller than the value favored by global fits~\cite{deSalas:2017kay} (however, note that this phase is not yet directly measured), and finally, (iv) the neutrino mass spectrum is $m_\nu \simeq \{3.54\times 10^{-3},9.55\times 10^{-3}, 4.94 \times 10^{-2}\}\, \mathrm{eV}$ for the light neutrinos and $M_N \simeq \{2.09\times 10^{10}, 3.88\times 10^{11},2.08\times 10^{12} \}\, \mathrm{GeV}$ for the right-handed neutrinos.

It is well known that inverted neutrino mass ordering is much more difficult to fit in SO(10) models than normal ordering (see e.g.~Refs.~\cite{Joshipura:2011nn,Dueck:2013gca}). Indeed, the best-fit for inverted ordering has $\chi^2 \simeq 918$. It is interesting to note that global fits to neutrino oscillation data are also disfavoring inverted ordering~\cite{deSalas:2017kay}. As noted in other studies, it was also not possible to accommodate thermal leptogenesis (we find that $\chi^ 2 \simeq 413$ in the case of normal ordering) via the decay of the lightest right-handed neutrino, due to constraints imposed by the SO(10) symmetry. Following Ref.~\cite{Giudice:2003jh}, we have calculated the leptogenesis prediction by solving the Boltzmann equations for the decay of the lightest right-handed neutrinos including washout effects (but not the flavor effects). However, a more precise treatment of leptogenesis may result in better fits, e.g., taking into account the contributions of the $\mathrm{SU}(2)_L$ triplet in the $\repb{126}_H$ \cite{Antusch:2004xy,Antusch:2007km} (which would also contribute to neutrino masses via type-II seesaw~\cite{Lazarides:1980nt,Mohapatra:1980yp,Wetterich:1981bx}), other scalars in the $\repb{126}_H$~\cite{Fong:2014gea}, or the strong thermal leptogenesis solution \cite{DiBari:2013qja,Chianese:2018rnq}.

\begin{figure}
\includegraphics[clip,width=0.45\textwidth]{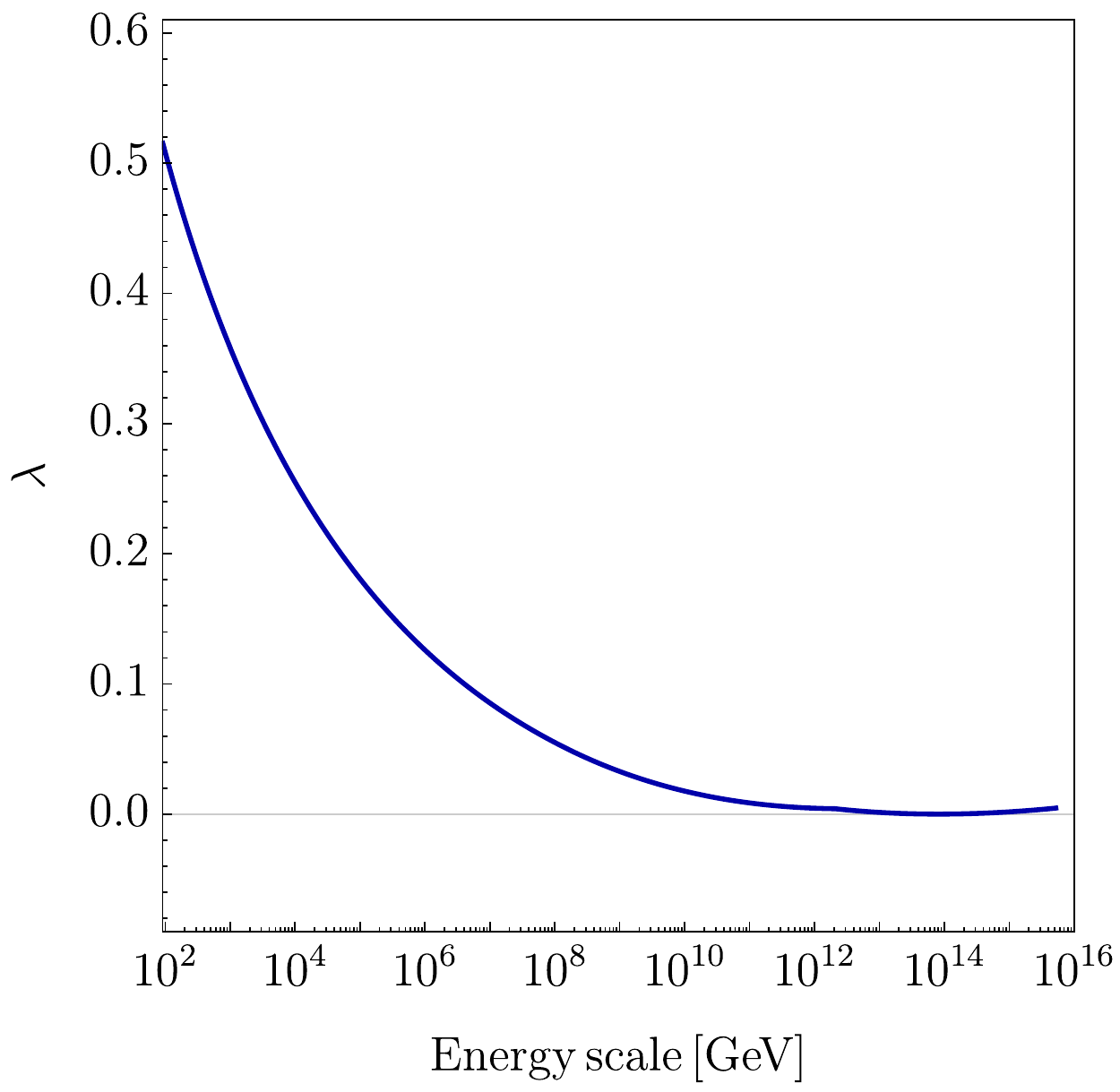}
\caption{\label{fig:lambda}Variation of the Higgs quartic coupling $\lambda$ with energy in our model. The kink at $2.08\times 10^{12}\,\mathrm{GeV}$ is due to the heaviest right-handed neutrino being integrated out.}
\end{figure}

Our model also allows for a solution to the instability problem of the electroweak vacuum. This is directly associated to the Higgs quartic coupling $\lambda$ (see e.g.~Ref.~\cite{Buttazzo:2013uya}) and can be solved by adding appropriate new physics not far from the TeV scale. In our case, this is readily provided by the color-octet scalar $S_2$, and to a lesser extent the other color-octet scalar $S_1$ since $S_2$ has the largest effect on the RG running of the gauge couplings. As noted earlier, the fits were performed with $\lambda(M_\mathrm{GUT})=0$. However, we   observed that this did not fully prevent $\lambda$ from becoming negative, albeit in a small energy region. To rectify this, we compensate by shifting $\lambda(M_\mathrm{GUT})$ up by a small amount ($0.005$ for the best-fit) such that it remains positive throughout the whole energy range between $M_Z$ and $M_\mathrm{GUT}$. We verify that this small shift has a negligible effect on the observables in the fit. In Fig.~\ref{fig:lambda}, we present the RG running of $\lambda$ in our model at one-loop order, taking into account all relevant contributions. Although the contributions of $S_{1}$ and $S_{2}$ are somewhat counter-balanced by that of the right-handed neutrinos, the total effect remains positive.

\section{Axion Dark Matter and Inflation}
\label{sec:axions}

Our model provides invisible axions as a solution to the dark matter problem via the Dine--Fischler--Srednicki--Zhitnitsky mechanism \cite{Zhitnitsky:1980tq,Dine:1981rt}. Since the $\mathrm{U}(1)_\mathrm{PQ}$ symmetry breaks at $M_\mathrm{GUT}$, the axion decay constant is $f_A \sim M_\mathrm{GUT} \approx 4.51\times 10^{15}$~GeV. For this value, the upper bound on the isocurvature  fluctuations \cite{Axenides:1983hj,Linde:1985yf,Linde:1991km,Linde:1984ti,Seckel:1985tj,Turner:1990uz} constrains the inflation energy to be smaller than about $6.8\times 10^8\, \mathrm{GeV}$ \cite{Akrami:2018odb}, implying that the PQ symmetry is broken before inflation and the correct abundance of axion dark matter is fixed anthropically by tuning the value of the misalignment angle to be about $3.9 \times 10^{-3}$.

\section{Summary and Conclusions}
\label{sec:conclusions}

Non-super\-symmetric GUTs based on $\mathrm{SO}(10)$ gauge symmetry provide a promising framework for new physics. We have investigated a minimal non-super\-symmetric $\mathrm{SO}(10)\times \mathrm{U}(1)_\mathrm{PQ}$ model, which breaks directly to the SM. Gauge coupling unification is achieved by splitting a representation contributing to the breaking of $\mathrm{SO}(10)$, namely the $\mathbf{210}_H$, such that two color-octet scalar representations have intermediate masses. We have determined the proton lifetime to be below $4.5\times 10^{34}$ years, in reach of the sensitivity of the proposed Hyper-Kamiokande experiment. In particular, we have observed that the non-observation of proton decay at Hyper-Kamiokande after five years, combined with the lower bound on the mass from LHC searches, would strongly disfavor our model in its minimal realization. Furthermore, the two color-octet scalars help stabilize the electroweak vacuum. We have performed numerical fits to the para\-meters of the model and found a reasonable agreement with data in the case of normal neutrino mass ordering. We have predicted the unknown neutrino parameters, and in particular, the leptonic CP-violating phase $\delta_\mathrm{CP}^\ell \simeq 0.187$. The PQ symmetry solves the strong CP problem and provides a dark matter candidate in the form of axions produced via the misalignment mechanism in the anthropic window. Although we have not discussed details of the inflationary scenario in this model, we have concluded that its scale should be below $6.8\times 10^8\,\mathrm{GeV}$. 
It would be worthwhile to investigate baryogenesis and inflation in our model in more detail to have a model addressing all the shortcomings of the SM in a minimal SO(10) GUT (as in the model based on $\mathrm{SU}(5) \times \mathrm{U}(1)_\mathrm{PQ}$ presented in Ref.~\cite{Boucenna:2017fna}).

\begin{acknowledgments}
S.M.B.~thanks the ``Roland Gustafssons Stiftelse f{\"o}r teoretisk fysik'' for partial financial support. T.O.~acknowledges support by the Swedish Research Council (Vetenskapsr\r{a}det) through contract No.~2017-03934 and the KTH Royal Institute of Technology for a sabbatical period at the University of Iceland. M.P.~thanks ``Stiftelsen Olle Engkvist Byggm{\"a}stare'' for financial support through contract No. 2017/85 (179) as well as ``Roland Gustafssons Stiftelse f{\"o}r teoretisk fysik''. Numerical computations were performed on resources provided by the Swedish National Infrastructure for Computing (SNIC) at PDC Center for High Performance Computing (PDC-HPC) at KTH Royal Institute of Technology in Stockholm, Sweden under project numbers PDC-2018-49 and SNIC 2018/3-559.
\end{acknowledgments}

\bibliographystyle{apsrev4-1}
\bibliography{refs.bib}

\end{document}